# Graphene-Based Non-Boolean Logic Circuits

Guanxiong Liu, Sonia Ahsan, Alexander G. Khitun, Roger K. Lake and
Alexander A. Balandin[*]

Department of Electrical Engineering, University of California – Riverside,
Riverside, California 92521 USA

## Abstract

Graphene revealed a number of unique properties beneficial for electronics. However,
graphene does not have an energy band-gap, which presents a serious hurdle for its
applications in digital logic gates. The efforts to induce a band-gap in graphene via
quantum confinement or surface functionalization have not resulted in a breakthrough.
Here we show that the *negative* differential resistance experimentally observed in
graphene field-effect transistors of "conventional" design allows for construction of
viable *non-Boolean* computational architectures with the gap-less graphene. The
negative differential resistance – observed under certain biasing schemes – is an
intrinsic property of graphene resulting from its symmetric band structure. Our
atomistic modeling shows that the negative differential resistance appears not only in
the drift-diffusion regime but also in the ballistic regime at the nanometer-scale –
although the physics changes. The obtained results present a conceptual change in
graphene research and indicate an alternative route for graphene's applications in
information processing.

---

[*] Corresponding author (AAB): balandin@ee.ucr.edu

Modern digital logic is based on Boolean algebra implemented in semiconductor switch-based circuits [1]. For more than half-century, downscaling of silicon complementary-metal-oxide-semiconductor technology (CMOS) provided increasing performance of computer chips and enabled progress in information technologies. However, as the electronic industry leaders are working on the sub 10-nm technology node, it is widely expected that the downscaling of Si CMOS technology will not last much further beyond 2026 [2]. The problem of heat dissipation and physical limitations of silicon are expected to end the "era of silicon" computer chips, which enabled progress in information technologies. This fact motivates a search for alternative materials and computational paradigms that can, if not replace Si CMOS, then complement it in special-task information processing [3-4].

Since its first mechanical exfoliation [5] and discovery of its extraordinary high mobility at room temperature (RT) [6], graphene attracted attention as a potential candidate for future electronics. In addition to its high mobility, graphene reveals exceptional heat conduction properties [7], high saturation velocity [8], convenient planar geometry and capability for integration with virtually any substrate [9]. However, the absence of the energy band-gap, $E_G$, in graphene means that graphene device cannot be switched off resulting in the high leakage currents and prohibitive energy dissipation. A large number of research groups have attempted to solve this problem via application of an electric field [10, 11], quantum confinement of carriers in nanometer-scale ribbons [12], surface functionalization with various atoms [13, 14] and strain engineering [15, 16]. The outcome of these efforts was a modest band gap opening of few-hundred meV, which often came at the expense of strongly degraded electron mobility. Practical applications of graphene in digital circuits would require a band-gap on the order of 1 eV at room temperature (RT).

Here we describe a departure from the conventional approaches for graphene's electronic applications. We intentionally avoid any attempt to artificially induce an energy band, which would make graphene "more-silicon-like", and allow us to use Si CMOS architectures. In addition, we neither use tunneling effects [17, 18] in the device designs in order to keep the device structure and technological steps as simple as possible. The mechanism of the negative differential resistance (NDR) effect, which we experimentally observed in the large-size graphene devices operating in the drift-diffusion regime, is similar to that recently reported in Ref. [19]. However, we do not use high electrical fields to induce non-uniform doping to activate the NDR. Through the first-principle atomistic modeling we show that the NDR effect holds in the ballistic transport regime, which is characteristic for downscaled architectures. The proposed alternative computational paradigm makes use of the described NDR effect and can be effectively implemented with the *gap-less* graphene. Our graphene logic circuitry is based entirely on the intrinsic NDR effect in graphene and benefits from graphene's high electron mobility and thermal conductivity.

We start with the experimentally found conditions for observing the NDR effect in the dual-gate graphene field-effect transistors (G-FETs) and the means of controlling its strength. The devices for this study are fabricated from mechanically exfoliated graphene on a $Si/SiO_2$ substrate [20, 21]. Micro-Raman spectroscopy is used to select samples of single layer graphene (SLG) and bilayer graphene (BLG). The details of our micro-Raman procedures for graphene quality control were reported elsewhere [22, 23]. The source, drain, and gate regions made of Ti and Au are defined by the electron-beam lithography (EBL). The top-gate oxide is deposited using the two-layer method. The first layer is a thin film of evaporated Al, which is oxidized in air [24]. The second layer is grown by ALD. The heavily doped Si substrate acts as the back-gate. Figure 1 (a) shows a typical scanning electron microscopy (SEM) image of the dual-gate G-FET. The transfer characteristics of the BLG device under different back-gate voltages, $V_{BG}$, are shown in Figure 1 (b). The decreased Dirac point conductivity at large $V_{BG}$ indicates a transport gap opening in the BLG by the displacement filed [10, 11]. The transport gap induced by $V_{BG}$ in BLG is beneficial for NDR effect but it does not play a key role. In fact, as it will be shown later, the mechanism of the onset of NDR effect in our approach is the same for SLG and BLG.

In order to obtain NDR characteristics in the dual-gate G-FETs, which can be used for logic functions, we applied an unconventional biasing scheme. The conventional bias sets only one terminal with variable input, while all the other terminals are fixed at constant values. For example, in the source-drain current $I_{DS}$ versus the gate voltage $V_{GS}$ measurement, the source-drain voltage $V_{DS}$ and back-gate voltage $V_{BG}$ are fixed, while $I_{DS}$ is controlled by sweeping the top-gate voltage $V_{TG}$. Interestingly, the dual-gate G-FET reveals NDR when we sweep the $V_{DS}$ and $V_{TG}$ simultaneously. As $V_{DS}$ increases from zero while $V_{TG}$ scans across the Dirac point – the NDR effect occurs. The magnitude of the peak-to-valley ratio, $I_P/I_V$, can be tuned by the back-gate voltage.

Figure 1 (c) shows a clear *N*-shape NDR in the dual-gate G-FET at the $V_{BG}$=70V when $V_{DS}$ is varying from 0 to -0.1 V and $V_{TG}$ is changing from 0 to -6 V. To obtain this characteristic the $V_{DS}$ and $V_{TG}$ voltages were swept simultaneously following the same number of steps as $V_{BG}$ was fixed at certain value. The obtained current-voltage (I-V) curve shows a positive conductance in the regions I and III, and the negative conductance in the region II. Note that the magnitude of the peak current, $I_P$, is about a factor of 28 larger than that of the valley current, $I_V$. The differential conductance, $dI/dV$, has a negative peak value of -0.58 mS while the positive value is about 0.6 mS. By fixing $V_{BG}$ at a different voltage – while keeping the $V_{DS}$ and $V_{TG}$ sweep setting – the strength of the NDR effect can also be tuned. Figure 1 (d) indicates that $I_P/I_V$ decreases as $V_{BG}$ increases from -50 V to 70 V. The transition points of the conductance from the positive to the negative and from the negative to the positive shifts to smaller $V_{DS}$ and the NDR region shrinks with increasing $V_{BG}$. Figure 1 (e) and (f) show the transfer characteristics of SLG and the NDR effect as we sweep $V_{DS}$ and $V_{TG}$ simultaneously. It is clear from these data that even without any transport gap, induced by the displacement field, the NDR effect is still pronounced. The key for NDR activation is the symmetric band structure of graphene and its high carrier mobility. The transport gap induced in BLG device can help one to increase the peak-to-valley ratio (see *Methods* for details).

Qualitatively, the NDR effect in graphene can be understood from the contour map in Figure 2 (a). It shows the $V_{DS}$-$I_{DS}$ curves under different $V_{TG}$. In this plot, the x-axis is $V_{TG}$ sweeping from 0 to -6 V and y-axis is $V_{DS}$ sweeping from 0 to -1 V. Our biasing schematic of simultaneous sweeping of $V_{DS}$ and $V_{TG}$ is equivalent to drawing a diagonal line on this contour map. As long as the Dirac point falls within the range that $V_{TG}$ swept, the diagonal line will cross the region where the source-drain current $I_{DS}$ decreases with increasing $V_{DS}$. Figure 2 (b) shows the current profile along the diagonal line. One can see that NDR effect happens between the points B and C. The quantitative description of the effect within the drift-diffusion model of electric currents in graphene is given in the *Methods* section. As the source-drain bias $V_{DS}$ increases in the simultaneous sweep, shown in Figure 2(c) and (d), we find that the peak-to-valley ratio reduces. When the $V_{DS}$ and $V_{TG}$ voltages are swept within the same range from 0 to - 4V, shown in Figure 2 (e), the NDR effect becomes small. The reason is that as $V_{DS}$ increases, the carrier concentration of the graphene channel becomes dependent not only on the gate bias but also on the drain voltage [19]. As Figure 2 (f) shows, the transfer characteristics of the GFET are greatly broadened when $V_{DS}$ increase from -0.1 V to -4.9V. In this case, a much larger gate voltage is needed to change the graphene channel from *n* type to *p* type. The latter weakens the NDR effect.

Our biasing configuration is analogous to the diode connected metal-oxide-semiconductor field-effect-transistors (MOSFETs), which is widely used in modern integrated circuits. The diode connected MOSFETs, where the gate is connected with the drain, behaves similar to a diode so that the current starts to increase only when $V_{DS}$ and $V_{TG}$ are larger than the threshold voltage. In our approach, the onset of NDR is the result of interplay between the decreasing carrier concentration and the increasing electrical field along the graphene channel. The high carrier mobility of graphene and high resistance value at the Dirac point are essential factors for observing a pronounced peak-to-valley ratio (see the analysis at *Method*). The sweeping range of $V_{TG}$ and $V_{DS}$ is defined by the top-gate capacitance and the Dirac point position. The larger the top-gate capacitance the smaller the sweeping range of $V_{TG}$ is needed. Owing to the technological limitations of the gate oxide (consisting of ALD deposited 2 nm / 10 nm $AlO_x/HfO_2$) we have to use $V_{TG}$ within the range of values that are several times larger than those of $V_{DS}$. Due to the *n*-type doping nature of our devices the polarity was chosen negative for both $V_{TG}$ and $V_{DS}$. In principle, implementing the devices with a thinner gate oxide and higher dielectric constant [25] one can achieve a strong NDR effect within the 1-2 V range. Tying the gate and drain together one can readily transform a double gate G-FET into a three-terminal NDR device with the widely tunable peak position and peak-to-valley ratio.

The experimental device is large and operates in the drift-diffusion regime. However for practical applications one has to consider electron transport in downscaled computer architectures with the devices feature sizes on the order of a few nanometers. Here, we theoretically analyze a highly scaled version of the device that operates in the *ballistic*, quantum-capacitance limit, and we determine whether such an FET, in a diode-connected configuration, will have a current-voltage response exhibiting NDR. Transmission and current-voltage responses are calculated using an atomistic Huckel model within the nonequilibrium Green's function formalism (see *Methods* section for details). A schematic diagram of a single layer graphene FET is shown in Figure 3 (a). To investigate transport properties in the quantum capacitance regime, we consider a 3 nm gate oxide with a dielectric constant of 25. The calculated gate oxide capacitance ($C_G$) is 7.3 µF/cm². The device is in the quantum capacitance regime when $C_G > C_Q$ where $C_Q$ is the quantum capacitance of the channel [26]. In equilibrium, the source-to-drain potential profile is that of an npn structure in which the source and drain are n-type and at the same potential, and the channel is p-type. The built in potential ($V_{pn}$) between the source and the channel region as shown in Figure 3 (b) (inset) is $V_{pn}=2 \mu_s$ where $\mu_s$ is measured from the charge-neutral point of the source. The current-voltage response shown in Figure 3 (b) is calculated for a diode-connected, single-layer G-FET (SLGFET), i.e. the gate is shorted to the drain. The I-V response does exhibit NDR, and for a higher value of $\mu_s$, the peak to valley current ratio increases. The I-V response demonstrates NDR for an effective 22 nm channel operating in the ballistic limit and the quantum capacitance regime.

This regime is the opposite of the diffusive regime of the experimental device. Although the transport physics is qualitatively different, the physical mechanism governing the NDR is qualitatively the same. NDR results when the Dirac cone in the channel can be moved sufficiently fast with respect to the gate voltage in the drain. In a diode-connected G-FET in the quantum capacitance regime, this ratio is 1:1. The origin of the NDR behavior of the ballistic device can be described by the transmission curves shown in Figure 3 (c) and the corresponding band alignments shown in the insets. At low bias, the transmission is given by the red curve corresponding to the band-alignment shown in the left inset. The transmission is limited by the transition between the source conduction band and the channel valence band. Conservation of energy and momentum cause the transmission to be proportional to the area of the overlapping inverted triangles representing the electron and hole dispersions. The minimums in the red transmission curve correspond to the energies of the charge neutral points in the source and the channel. The current is proportional to the area under the transmission curve between the source Fermi level ($\mu_s$) and the drain Fermi level ($\mu_{d1}$) shown on the transmission plot. As the bias turns on, this area initially increases and the current increases.

As the bias continues to increase, the charge-neutral point of the channel is pulled down into the energy window between the source and drain Fermi levels as shown in the right inset of Figure 3 (c) resulting in the blue transmission curve. The two minimums in the transmission again correspond to the charge neutral points that have now been brought closer together in energy. The transmission regions labeled 'D' and 'C' result from unipolar transport between the source and channel, hole-hole and electron-electron, respectively. The region labeled 'B' lying between the two charge-neutral points results from interband transport between the source conduction band and the channel valence band. The minimum in transmission at negative energies outside of the domain of the graph results from the charge-neutral point of the drain. At this bias, even though the difference between the source and drain Fermi levels, $\mu_s$ and $\mu_{d2}$, has increased, the area under the transmission curve is a minimum, resulting in the current minimum and NDR. The current-voltage response of the diode-connected bilayer G-FET (BLGFET) is similar to that of the SLGFET. A dual, 3-nm, high-K, top and bottom gate are required to keep the BLGFET in the quantum capacitance limit. The gates are shorted, so that the two layers of the bilayer are at equal potential. A comparison of the SLGFET and the BLGFET with the same Fermi levels and built in potentials is shown in Figure 3 (d). The peak-to-valley current ratio of 2.0 for the BLGFET is slightly greater than the PVCR of 1.8 for the SLGFET. The analysis of the transmission for the BLGFET is similar to that of the SLGFET. Although the density of states is finite at the charge neutral point, it is still a minimum, and the transmission curves look qualitatively the same as in Figure 3 (c).

The NDR effect, experimentally observed in the drift-diffusion transport regime and theoretically predicted in the ballistic transport regime, allows one to use *pristine* graphene in information processing. In order to fully utilize graphene's unique properties we envisioned alternative logic circuits based on the diode-connected G-FETs. The ability to control NDR with $V_{BG}$ provides an additional degree of freedom for logic circuit design. The operation of the proposed circuits is illustrated with numerical simulations using analytical $I_{DS}$ vs. $V_{DS}$ curves extrapolated from the experimental and the theoretical data (see Figure 4 (a)). Without the loss of generality, we assumed that the back-gate capacitance, $C_{BG}$, is one half of the top-gate capacitance, $C_{TG}$, ($C_{BG}$=0.5$C_{TG}$) and introduced a linear shift for the $V_{TGs}^0$ as follows: $V_{TGs}^0 =1-0.5V_{BGs}$. The current is shown in normalized units of $I_0 = (W/2L)\mu e n_0 V_{BGs}^0$. There are two general trends in the *I-V* response to $V_{BG}$ to capture: (i) the NDR region shifts to the left with increasing $V_{BG}$ and (ii) the NDR region shrinks with increasing $V_{BG}$.

As a building block for the graphene logic gate construction we consider a circuit combining two G-FETs connected in series as shown in Figure 4 (b). It is a four terminal device, where the two back gates serve as the input terminals, one control input with $V_{DD}$, and the common top gate serves as the output terminal. The output voltage depends on three parameters: $V_{DD}$, $V_{BG1}$, and $V_{BG2}$, which can be controlled independently. There may be one or two stable outputs depending on the combination of the control voltages. The plots in Figure 4 (c) illustrate the possible scenarios for two *I-V* curves intersection leading to the single- or bi-stable output. In Figure 4 (d) we show a color map of the possible output voltages depending on the two back-gate voltages $V_{BG1}$ and $V_{BG2}$ at a fixed $V_{DD}$. The red color depicts the region in the 2D space where the output has two stable values. The dark blue and the light-blue color depict the regions of the single-value output (e.g. dark-blue color shows the "low" output $V_{OUT}<0.5V_{DD}$, and the light-blue color shows the "high" output $V_{OUT}>0.5V_{DD}$).

The devices shown in Figure 4 (b) can be used as a building block for a variety of logic gates. The regions with the single-valued output can be used for Boolean logic gates, while the bi-valued regions are of great promise for application in a non-Boolean logic, e.g. non-linear networks. For example, NAND and NOT logic gates can be realized by using the same circuit comprising two G-FETs connected in series as illustrated in Figure 4 (e-f). In order to build the inverter, the back gates of the two transistor diodes should receive the same input voltage $V_{BG1}=V_{BG2}$. Then, it is possible to find the $V_{DD}$ value leading to the inversion function. The results of numerical modeling in Figure 4 (e) show the input-output voltage dependence at $V_{DD}=3V_0$. The low input voltage corresponds to the high output, and vice versa, which is equivalent to logic NOT gate. The gain of the considered circuit, $\varDelta V_{OUT}/\varDelta V_{IN}$, depends on the strength of the back gate modulation α as well as on the peak-to-value ratio of the NDR region. NAND gate can be also realized by using the same circuit with a proper choice of $V_{DD}$ and input voltages. The process of finding the right input voltages is illustrated in Figure 4 (f). As an input data, we took the results presented in Figure 4 (d). In order to build a NAND gate, we need to find the region in the map where the low-input voltages (logic 0) correspond to the high output (logic 1). Input voltages of $1.7V_0$ and $4.5V_0$ satisfy such a condition at $V_{DD}=3V_0$. All other Boolean logic gates can be constructed via different variations of NAND gate.

However, the potential of NDR characteristics of G-FETs can be more fully realized in building the non-Boolean logic architectures. The concept of the non-linear network based on the devices with RTD is a well-known example of a non-Boolean approach [27]. To date this approach was limited to the tunneling diodes, which are the two-terminal devices. Utilization of the diode-connected G-FETs offers a three-terminal device with NDR. The latter allows one to build ultra-fast non-linear networks with enhanced logic capabilities. In Figure 5 (a) we show a circuit, which combines three layers – stages – of G-FETs, where each layer consists of two G-FETs connected in series. Each stage is biased by a separate $V_{DD}$, with the value that can vary from stage to stage. The input voltages are applied to the back gates of the transistors. The top gates of each stage are connected to the one of the back gates of the next stage. The main idea is to make use of the bi-stable outputs provided by each stage and to build a multi-valued logic unit. The results of numerical modeling presented in Figure 5 (b-c) illustrate the output values (black markers) after each stage as well as the ensemble of the output values after the last stage. In this example, all inputs are chosen to have either $0.5V_0$ or $1.9V_0$ (this combination leads to the bi-stable output at $V_{DD}$=3.0V as shown in Figure 4 (c)). Thus, the output of the Stage 1 may have five possible stable values for the four possible input combinations. The number of possible output values depends on the inputs as well on the Stage bias voltage $V_{DD}$. The results of our numerical modeling show three evolution trees for different combinations of the stage $V_{DD}$. We intentionally use the input and $V_{DD}$ values leading to the increased number of the output states.

At some point, the patterns shown in Figures 5 (b-c) resemble the operation of the Neumann cellular automata [28], where the existence or absence of the stable output in the certain value interval is analogous to the logic 0 or 1. For example, one can imagine the whole space of possible output voltages to be divided into the cells, e.g. of $0.5V_0$ width. The presence of a stable output in the given voltage interval can be interpreted as a logic 0, and the absence of the stable output can be assigned to logic 1. One can consider this type of the multi-valued network as the "voltage-space cellular automata", where an individual cell is not related to a real circuit or structure. Though there is no physical object related to an individual cell, the logic output can be easily recognized by measuring the output voltages. The presented network built of G-FETs can be modified in a number of ways, e.g. by increasing the number of transistors per stage, introducing a time-varying bias voltage, $V_{DD}(t)$, or increasing the number of interconnects among the stages. The proposed ultra-fast non-Boolean logic circuits implemented with G-FETs connected to reveal NDR characteristics can be used to construct a new type of cellular automata particularly suitable for special task data processing such as image recognition, data encryption, and database search.

Special task data processing logic circuits is another promising direction for G-FET implementation. It would be of great practical benefit to develop graphene-based analog logic circuits able to complement of complementary metal-oxide semiconductor (CMOS) technology in doing specific operations, which require enormous resources for the conventional digital counterparts. Pattern matching is one of the examples, which is widely-used for database search, spell checking and signal processing tasks. The essence of the pattern matching operation is the checking if the stream of input data matches a reference one. The main challenge for this application is to perform high throughput operation to match the speed of the gigabit network. The inevitable development of 100 Gbps-scale data networks would make real time network intrusion detection impossible [29] even using the most optimistic assumptions for scaling CMOS [30]. That is one of the cases where unique properties of graphene may be utilized to complement the existing technology.

The schematics of the pattern matching circuit built of G-FETs are shown in Figure 5 (d). The circuit consists of a number of similar cells connected in series, where the elementary cell comprises three G-FETs arranged in a two-stage circuit. The input data are applied to the two inputs of the first stage G-FETs. The input voltages $V_1$ and $V_2$ represent two logic states 0 and 1, respectively. One of these voltages corresponds to the input data stream and the other corresponds to the reference data. The values of the input voltages are selected to provide the same voltage output of the first stage if and only if $V_1 = V_2$, which corresponds to the logic states 00 or 11. The output voltage of the first stage is then applied to the back gate of the second stage transistor. The output voltage corresponding to 00 and 11 states is matched to the Dirac point providing minimum conductivity of the second stage transistor as illustrated in Figure 5 (e). Overall, the elementary cell acts as a XOR gate providing minimum current for 00 and 11 inputs. The complete circuit consists of a number of XOR gates connected in series through the second stage transistors. The current flowing through the upper transistors decreases with the decrease of the Hamming distance among the input and reference data strings. It has absolute minimum in the case of the perfect match where zeros and ones of the input data matches zeros and ones of the reference data (see Figure 5 (f)).

The graphene-based pattern matching circuit shown in Figure 5 (d) has several important advantages in terms of area, speed and overall functional throughput over the existing circuits. One hand, the elementary XOR gate requires only three G-FETs (minimum 8 transistors in CMOS), where the area per graphene transistor can be as small as 10 nm ×40 nm [31]. All XOR gates are connected in series to the common sensing line allowing for parallel data read-in. On the other hand, the operation frequency of the graphene transistor can be as high as 427 GHz [32]. The maximum pattern matching throughput defined as $N_{bits} f_{max} / A_{cell}$ may exceed $10^{22}$ bits/s/cm$^2$, which is several orders of magnitude higher than for any reported or even projected scaled circuits [30]. This example illustrates the possibility of building special task analog logic circuits based on graphene devices, which can significantly outperform CMOS in one specific application.

In conclusion, we demonstrated that the negative differential resistance experimentally observed in graphene field-effect transistors allows for construction of viable non-Boolean computational architectures. The proposed approach overcomes the absence of the energy band gap in graphene. The negative differential resistance appears not only in the large scale graphene transistors but also in the downscaled devices operating in the ballistic transport regime. Our results may lead to a conceptual change in graphene research proving an alternative route for graphene's applications in information processing.

**METHODS**

**Drift-Diffusion Model of Electron Transport in NDR Regime without Tunneling**

The experimentally observed NDR effect can be explained in relatively simple terms using the drift-diffusion model for current conduction in graphene. We define $V_{DS}$ and $V_{TG}$ to sweep zero simultaneously with a different step size M, $V_{TG} = MV_{DS} = V$. Since we observed that the NDR effects happens close to the Dirac point where the $n_s$ is roughly proportional to $V_{TG}$, we can write that $n_{TG} = C_{TG}(-V_{TG} + V_{TG}^0)/e$, where $e$ is the elementary charge, $C_{TG}$ is the capacitance of the top-gate. We choose here the $p$-type branch of graphene since this is the region where NDR appears. Thus, we use $(-V_{TG} + V_{TG}^0)$, where $V_{TG}^0$ is the top-gate voltage at the Dirac point under a certain back-gate bias. We can write for the current $I = \frac{W}{L}[\sigma_S + \sigma_0]V_{DS}$, where $\sigma_S = \mu e n_{TG}$ is the conductivity controlled by the gate and $\sigma_0$ is the conductivity at the Dirac point. Thus, one arrives with the following equation:

$$I = \frac{W}{L}[\mu C_{TG}(-V_{TG} + V_{TG}^0) + \sigma_0]V_{DS}$$
$$= \frac{W}{L}[\mu C_{TG}(-V + V_{TG}^0) + \sigma_0]\frac{V}{M}$$

Taking the derivative of above equation and setting $\frac{\partial I}{\partial V} = 0$, we can find the peak value of the current achieved at $V_{peak} = \frac{1}{2}(\frac{\sigma_0}{\mu C_{TG}} + V_{TG}^0)$. The found peak current value is $I_{peak} = \frac{1}{4}\frac{W}{L}\frac{1}{\mu C_{TG}}(\sigma_0 + \mu C_{TG}V_{TG}^0)^2$. The valley current value, $I_{valley} = \frac{W}{L}\sigma_0 V_{TG}^0$, is reached at $V_{valley} = V_{TG}^0$. The peak-to-valley ratio is $\frac{I_{peak}}{I_{valley}} = \frac{1}{4}[\frac{\sigma_0}{\mu C_{TG}V_{TG}^0} + \frac{\mu C_{TG}V_{TG}^0}{\sigma_0} + 2]$. Plugging in the common values for our dual gate graphene devices, $\sigma_0 = 1/6k\Omega$, $\mu = 1000cm^2/Vs$, $C_{TG} = 0.94\mu F/cm^2$ for ~12 nm AlO$_x$/HfO$_2$ oxide stack and $V_{TG}^0 = -2V$ (tunable by back-gate voltage), we find that the absolute value of $\frac{\mu C_{TG} V_{TG}^0}{\sigma_0}$ is much larger than 1, so the

$$\frac{I_{peak}}{I_{valley}} \cong \frac{1}{4} [\frac{\mu C_{TG} V_{TG}^0}{\sigma_0} + 2].$$

From this equation, we can see that the higher mobility $\mu$, larger gate capacitance $C_{TG}$, the Dirac point far from a zero bias and a lower Dirac conductance $\sigma_0$ will be beneficial for increasing the peak-to-valley ratio $\frac{I_{peak}}{I_{valley}}$ of the NDR effect in the graphene devices. The width of the NDR region is determined by the difference between $V_{peak}$ and $V_{valley}$, $V_{valley} - V_{peak} = \frac{1}{2}(V_{tg}^0 - \frac{\sigma_0}{\mu C_{TG}})$ so that the requirement for appearance of the NDR effect is $V_{tg}^0 > \frac{\sigma_0}{\mu C_{TG}}$. Note that the $V_{TG}^0$ for SLG, and $V_{TG}^0$ and $\sigma_0$ in BLG are tunable by the back-gate voltage and the NDR effect in G-FET is tunable by the back-gate voltage.

**Atomistic Theory of Electron Transport**

A representative simulated schematic diagram of a SLGFET is shown in Fig. 3(a). The device consists of single layer graphene sheet as a conducting channel. The total channel length between the two leads is taken as 30nm for both the SLGFET and the BLGFET. For the BLGFET, two single-layer graphene sheets are stacked in AB alignment with an experimental separation distance of 3.35 Å.

Our atomistic model uses a Huckel Hamiltonian with a $p_z$ orbital basis. These atomic orbitals are approximated with Slater Type Orbitals [33]. The matrix elements of the Huckel Hamiltonian (H) are then described by the following equations [34-35]: $H_{i,i} = -V_i$ and $H_{i,j} = \frac{c}{2} S_{i,j} \left( H_{i,i} + H_{j,j} \right) \ \left( i \neq j \right)$. The diagonal elements of the Hamiltonian are approximated with the $p_z$ orbital ionization energies ($V_i$). The overlap matrix is $S_{i,j} = \langle i | j \rangle$ where $| j \rangle$ is a $p_z$ orbital on atom $j$. The off-diagonal elements are proportional to the overlap where the constant is taken as c =2.8 [33]. The matrix elements of the channel potential ($V$) are calculated as $\langle i | V | j \rangle = S_{i,j} [V(r_i) + V(r_j)] / 2$.

The device Hamiltonian, overlap matrix and the device-to-lead coupling matrices are used in the NEGF algorithm [36] to calculate the surface self-energies, Green's function and finally the transport characteristics of G-FETs. Our study addresses the ballistic transport through the channel. Throughout this work, the calculation is for room temperature. .To incorporate the bias voltage, we assume a constant shift of energy in the channel under the gate region. The potential changes linearly over a distance of 4 nm between the source and channel and between the channel and drain giving an effective channel length of 22 nm.

*Acknowledgements*

This work was supported, in part, by the Semiconductor Research Corporation (SRC) and Defense Advanced Research Project Agency (DARPA) through STARnet Center for Function Accelerated nanoMaterial Engineering (FAME). AAB and RKL also acknowledge funding from the National Science Foundation (NSF) and SRC Nanoelectronic Research Initiative (NRI) for the project 2204.001: Charge-Density-Wave Computational Fabric: New State Variables and Alternative Material Implementation (NSF ECCS-1124733) as a part of the Nanoelectronics for 2020 and beyond (NEB-2020) program. AAB also acknowledges funding from NSF for the project Graphene Circuits for Analog, Mixed-Signal, and RF Applications (NSF CCF-1217382).

## References

[1] Jaeger, R.C.; Blalock, T. N. *Microelectronic Circuit Design*; McGraw-Hill, 1997

[2] The International Technology Roadmap for Semiconductors, 2012

[3] Bourianoff, G., Brewer, J. E., Cavin, R., Hutchby, J. A. & Zhirnov, *V. Computer* 2008, 41, 38 − 46

[4] Zhirnov, V.V., Cavin, R.K, *J. Nanoelectron. Optoelectron.* 2006, 1, 52 − 60

[5] Novoselov, K. S.; Geim, A. K.; Morozov, S. V.; Jiang, D.; Zhang, Y.; Dubonos, S. V.; Grigorieva, I. V.; Firsov, A. A. *Science* 2004, 306, 666-669

[6] Blotin, K. I.; Sikes, K. J.; Jiang, Z.; Klima, M.; Fudenberg, G.; Hone, J.; Kim, P.; Stormer, H.L. *Solid State Commun.* 2008, 146, 351-355

[7] Balandin, A. A. *Nat. Mater.* 2011, 10, 569−581

[8] Schwiers, F. *Nat. Nanotechnol.* 2010, 5, 487

[9] Lin, Y.-M.; Jenkins, K. A.; Valdes-Garcia, A.; Small, J. P.; Farmer, D. B.; Avouris, P. *Nano Lett.* 2008, 9, 422-436

[10] Xia, F.; Farmer, D. B.; Lin, Y.-M.; Avouris, P. *Nano Lett.* 2010, 10, 715−718

[11] Zhang, Y.; Tang, T.; Girit, C.; Hao, Z.; Martin, M. C.; Zettl, A.; Crommie, M. F.; Shen, Y. R.; Wang, F. *Nature* 2009, 459, 820-823

[12] Han, M.Y.; Ozyilmaz, B.;Zhang, Y.;Kim, P. *Phys. Rev. Lett.* 2007, 98, 206805-8

[13] Zhang,W.; Lin,C.-T.; Liu, K.-K.; Tite, T.; Su,C.-Y.; Chang, C.-H.; Lee,Y.-H.;Chu, C.-W.; Wei,K.-H.; Kuo,J.-L; Li, L.-J. *ACS Nano* 2011, 5, 7517−7524

[14] Szafranek, B. N.; Schall, D.; Otto, M.; Neumaier, D.; Kurz, H. *Nano Lett.* 2011, 11, 2640−2643

[15] Ni, Z. H.; Yu, T.; Lu, Y. H.; Wang, Y. Y.; Feng, Y. P.; Shen, Z. X. *ACS Nano* 2008, 2, 2301−2305

[16] Choi, S.-M.; Jhi, S.-H.; Son, Y.-W. *Nano Lett.* 2010, 10, 3486 −3489

[17] Song, Y, ; Wu, H.-C. ; Guo, Y. *Appl. Phys. Lett.* 2013,102, 093118-093122

[18] Nguyen, V. H.; Niquet, Y.M.; Dollfus, P. *Semicond. Sci. Technol.* 2012, 27, 105018-105024

[19] Wu, Y.; Farmer, D. B.; Zhu, W.; Han, S.; Dimitrakopoulos, C. D.; Bol, A. A.; Avouris, P.; Lin, Y. *ACS Nano* 2012, 6, 2610− 2616

[20] Liu, G.; Stillman, W.; Rumyantsev, S.; Shao, Q.; Shur, M.; Balandin, A. A. *Appl. Phys. Lett.* 2009, 95, 033103-033105

[21] Yang, X.; Liu, G.; Balandin, A. A.; Mohanram, K. *ACS Nano* 2010, 4, 5532−5538

[22] Calizo, I.; Miao, F.; Bao, W.; Lau, C. N.; Balandin, A. A. *Appl. Phys. Lett.* 2007, 91, 071913-071915

[23] Calizo, I.; Bao, W.; Miao, F.; Lau, C. N.; Balandin, A. A. *Appl. Phys. Lett.* 2007, 91, 201904-01906

[24] Kim, S.; Nah, J.; Jo, I.; Shahrjerdi, D.; Colombo, L.; Yao, Z.; Tutuc, E.; Banerjee, S. K. *Appl. Phys. Lett.* 2009, 94, 062107-062109

[25] Dean, C. R.; Young, A. F.; Meric, I.; Lee, C.; Wang, L.; Sorgenfrei, S.; Watanabe, K.; Taniguchi, T.; Kim, P.; Shepard, K. L.; Hone, J. *Nat. Nanotechnol.* 2010, 82, 722− 726

[26] Rahman, A.; Guo, J.; Datta, S.; Lundstrom, M. S., *IEEE Trans. On Electron device* 2003, 50, 1853-1864

[27] Chua, L. O.; Yang, L. *IEEE Trans. on Circuits & Systems* 1988, 35, 1257-1272

[28] Neumann, J. *Theory of Self-Reproducing Automata*. Univ. Illinois Press: Urbana, IL, 1966

[29] Cho, Y. H.; Mangione-Smith, W. H. *ACM Trans. on Embedded Computing Systems* 2008, 7, 21:1-21:26

[30] Strukov, D. B. *Nanotechnology* 2011, 2, 9-12

[31] Liu, G.; Wu, Y.; Lin, Y.-M.; Farmer, D. B.; Ott, J. A.; Bruley, J.; Grill, A.; Avouris, P.; Pfeiffer, D.; Balandin, A. A. *ACS Nano* 2012, 6, 6786– 6792

[32] Cheng, R.; Bai, J.; Liao, L.; Zhou, H.; Chen, Y.; Liu, L.; Lin, Y.-C.; Jiang, S.; Huang, Y.; Duan, X. *Proc. Natl. Acad. Sci. U.S.A.* 2012, 109, 11588– 11592

[33] Mulliken, R. S.; Rieke, C. A.; Orlo, D. and Orlo, H., J. Chem. Phys. 1949, 17, 1248-1267

[34] Raza, H. and Kan, E. C., *J. Comput. Electron.* 2008, 7, 372-375

[35] Kienle, D.; Cerda, J. I. and Ghosh, A. W., J. Appl. Phys. 2006, 100, 043714-043722

[36] Datta, S. *Quantum Transport Atom to Transistor*; Cambridge University Press: Cambridge, 2005

**FIGURE CAPTIONS**

**Figure** 1: Experimentally observed negative differential resistance characteristics in graphene devices. (a) SEM of top-view SEM of a typical dual-gate graphene device. Gold color is the source/drain, pink color is the top gate and the blue color underneath is graphene flake. The gate and graphene channel is separated by a two-layer of $AlO_x$ and $HfO_2$ oxide stack. The scale bar is 1μm. (b) The transfer characteristics of BLG device under different back-gate voltage. The increased resistance at large back-gate voltage indicated band gap opening by perpendicular electric field. The inset shows the Dirac point shift as the back-gate voltage changes. (c) NDR effect happens on GFET as biasing were set such that the $V_{DS}$ ranging from 0 to -0.1 V, and $V_{TG}$ ranging from 0V to -6V, and $V_{BG}$=70V. The calculated dynamic conductance has the maximum negative value of -0.58 mS and maximum positive value of 0.6 mS. (d) Tunable NDR effects by changing the $V_{BG}$ from -50V to 70V. The $I_P/I_V$ increased as the $V_{BG}$ increased, and the negative resistance region also expands. (e) The transfer characteristics of SLG device under different back-gate voltage. (f) NDR effect in the same device. The data were obtained for $V_{DS}$ ranging from 0 to -1 V and $V_{TG}$ ranging from 0V to -4V under different $V_{BG}$.

**Figure 2:** Dependence of the negative differential resistance on the biasing conditions. (a) Contour plot of the $I_{DS}$ of GFET under various biasing conditions that $V_{BG}$ =50V, $V_{DS}$ and $V_{TG}$ sweeps from 0 to -1V and 0 to -6V, respectively. The diagonal line represents our simultaneous sweeping setup. (b) The source-drain current profile on this diagonal line explains the NDR effect in our sweeping biasing condition. As we increase the range of $V_{DS}$ to 0 to -1 V (c) and 0 to -3 V (d), the NDR effect is still preserved, but the $I_P/I_V$ decreases. (e) shows the result of connecting the drain with the gate, and applying the bias $V_{DS}=V_{TG}$ from 0 to -4 V. The NDR effect becomes much weaker. (f) shows the transfer characteristics under small $V_{DS}$ =0.1V (dark blue) and large $V_{DS}$ =4.9 V (orange). The gate effect is much stronger at small $V_{DS}$ than in large

$V_{DS}$. The Drain voltage effect results in non-uniform potential distribution along the graphene channel and broadened transition region around charge neutrality point.

**Figure 3:** Atomistic theory of the negative differential resistance effect in graphene devices. (a) Schematic diagram of the biased Drain-Gate shorted SLGFET device with the contact surface self-energies. The region inside the vertical line is the channel region. (b) *I-V* characteristics for different Fermi energy keeping $V_{pn}=2\mu_s$ of Drain-Gate shorted SLGFET.(inset: flat band potential profile in SLGFET). (c) Transmission coefficients as a function of energy for low and high bias where $\mu_s$ is the fermi level of the source and $\mu_{d1}$ and $\mu_{d2}$ are the fermi levels of drain contact at low and high bias respectively. (Inset: Energy spectrum of drain-gate shorted SLGFET for low and high bias region). (d) Comparison of *I-V* characteristics for SLGFET and BLGFET. Current plotted for $\mu_s$ =0.5eV and a built in potential of 1eV.

**Figure 4:** Implementation of logic gates with graphene without the energy band-gap. (a) Approximate $I_D$-$V_{DS}$ characteristics of the G-FET under different $V_{BG}$. (b) Schematics of the circuit comprising two graphene G-FETs. (c) Results of numerical modeling illustrating possible combinations of the input voltages leading to the single- value and bi- stable output. (d) Results of numerical simulations: the color map shows the output voltage as a function of the for different inputs at fixed $V_{DD}$. The red and the blue regions depict the bi-stable and single-valued output, respectively. The dark-blue color show "low" output ($V_{OUT}$<0.5 $V_{DD}$), and the dark-blue color show "high" output ($V_{OUT}$ ≥0.5 $V_{DD}$). (e) The results of numerical simulations illustrating the inversion function: low input results in the high output and vice versa (Input-1 = Input 2). (f) Possible NAND gate function with the proper choice of the input voltages.

**Figure 5:** Non-Boolean information processing with graphene circuits. (a) Schematics of the multi-stage network consisting of G-FETs. The input voltages are applied to the back gates of the transistors. The top gates of each stage are connected to the one of the back gates of the next stage. (b-c) Results of numerical modeling illustrating the evolution trees for output voltage at different combinations of the stage $V_{DD}$. (d) Schematics of the pattern matching circuit built of G-FETs. An elementary cell consists of three G-FETs arranged in a two-stage circuit. The elementary cell acts as a XOR gate providing minimum current for 00 and 11 logic input. (e) Results of numerical simulations showing the conductance of the second stage G-FET at four possible input combinations. (f) Illustration of the circuit functionality: current flowing through the upper transistors as a function of the Hamming distance among the input and reference data strings. It has absolute minimum in the case of the perfect match where zeros and ones of the input data matches zeros and ones of the reference data.

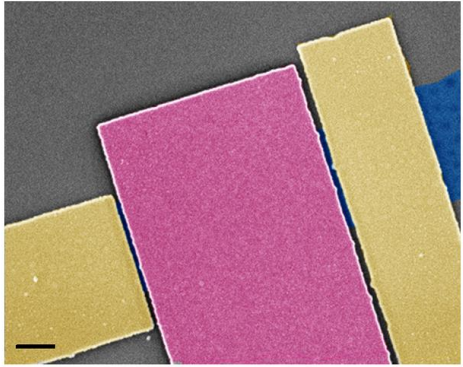

(a)

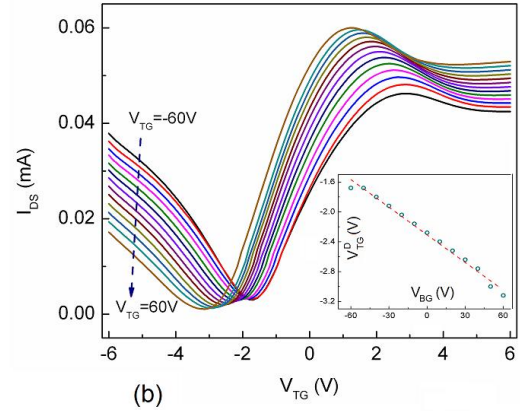

(b)

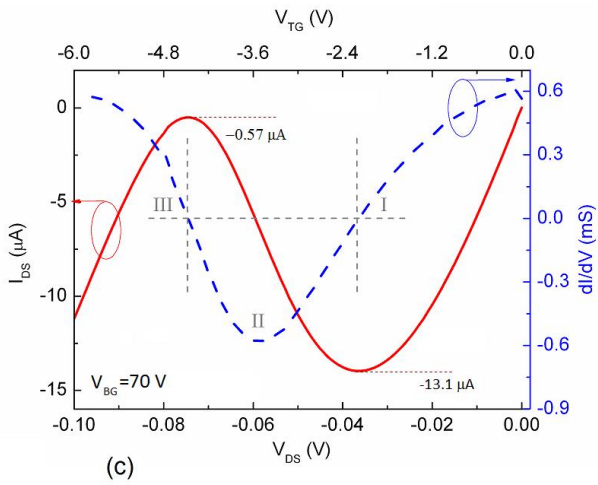

(c)

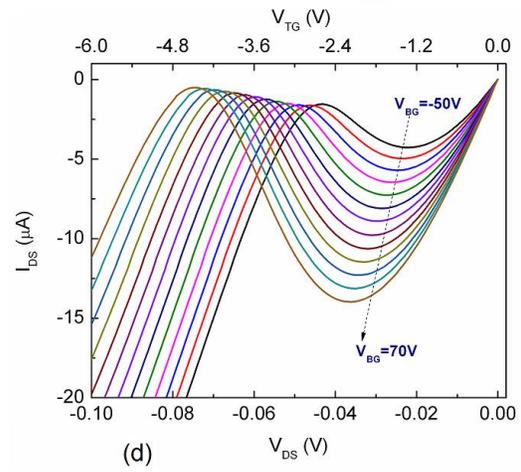

(d)

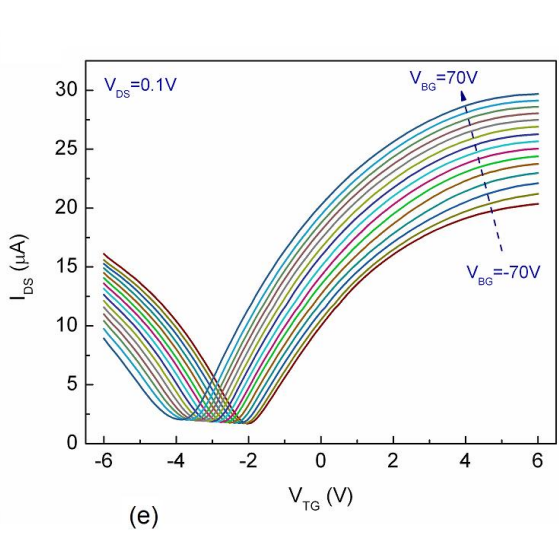

(e)

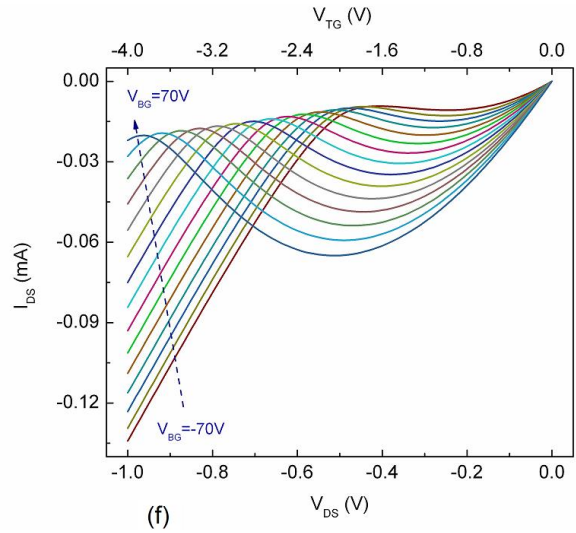

(f)

Figure 1 of 5

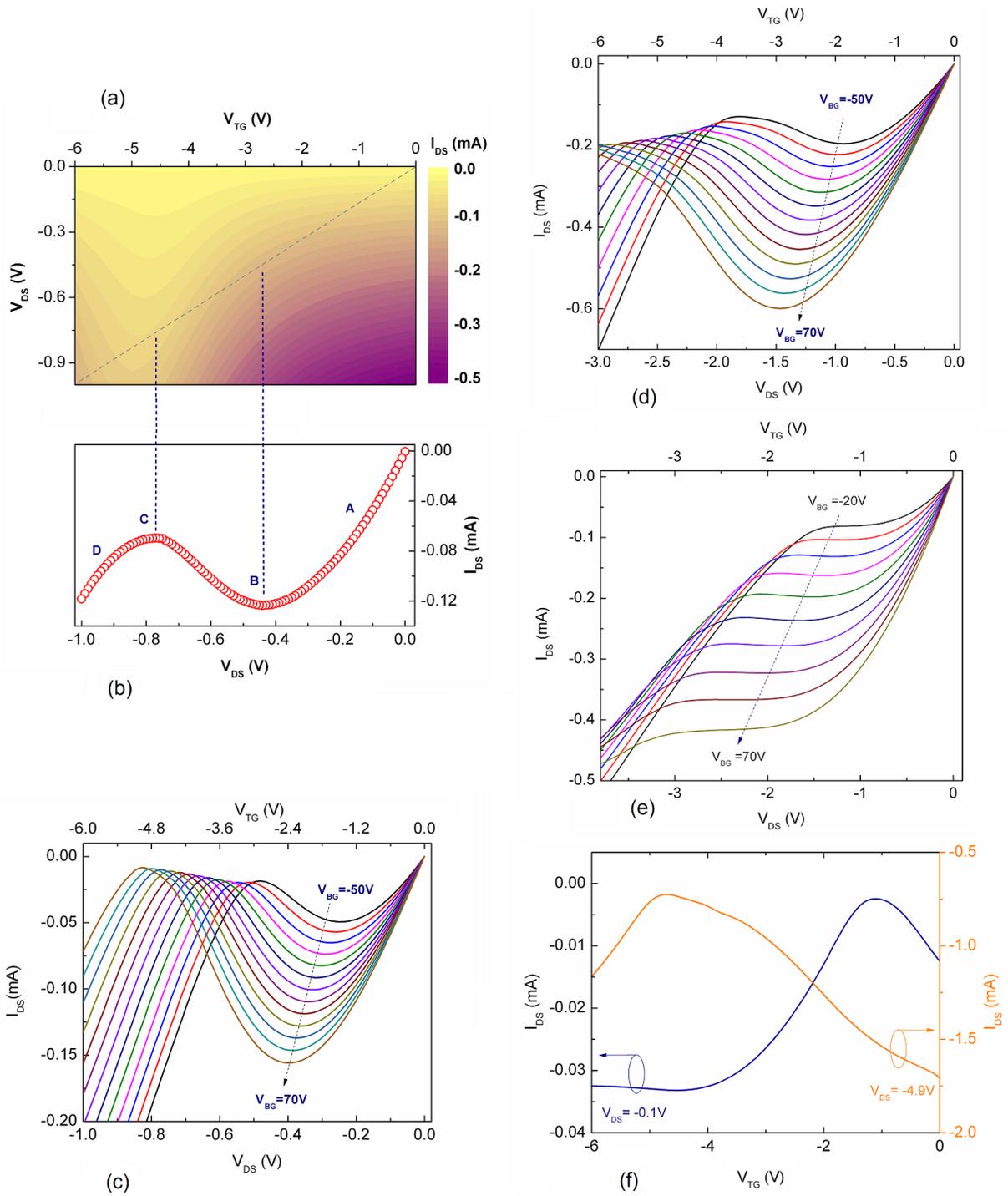

Figure 2 of 5

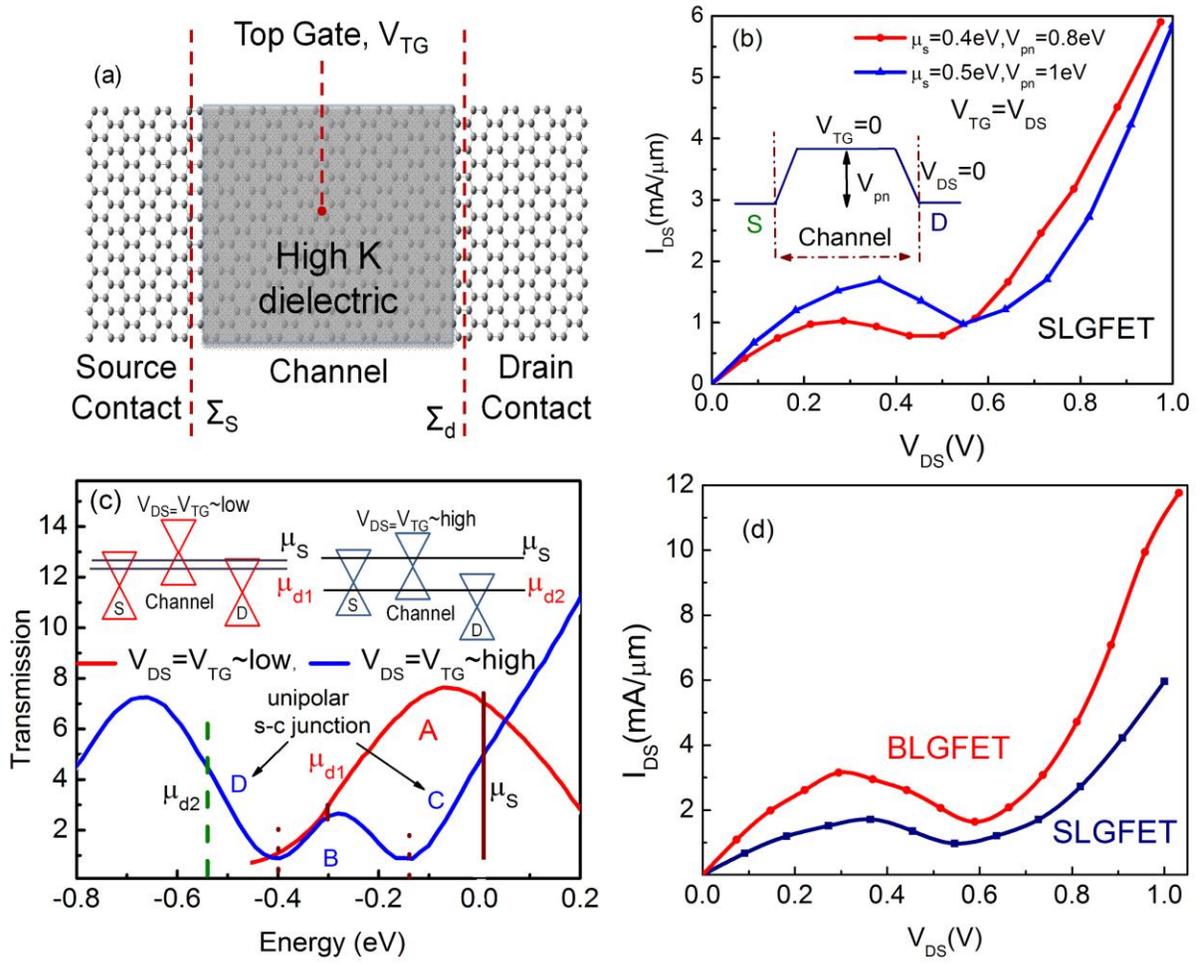

Figure 3 of 5

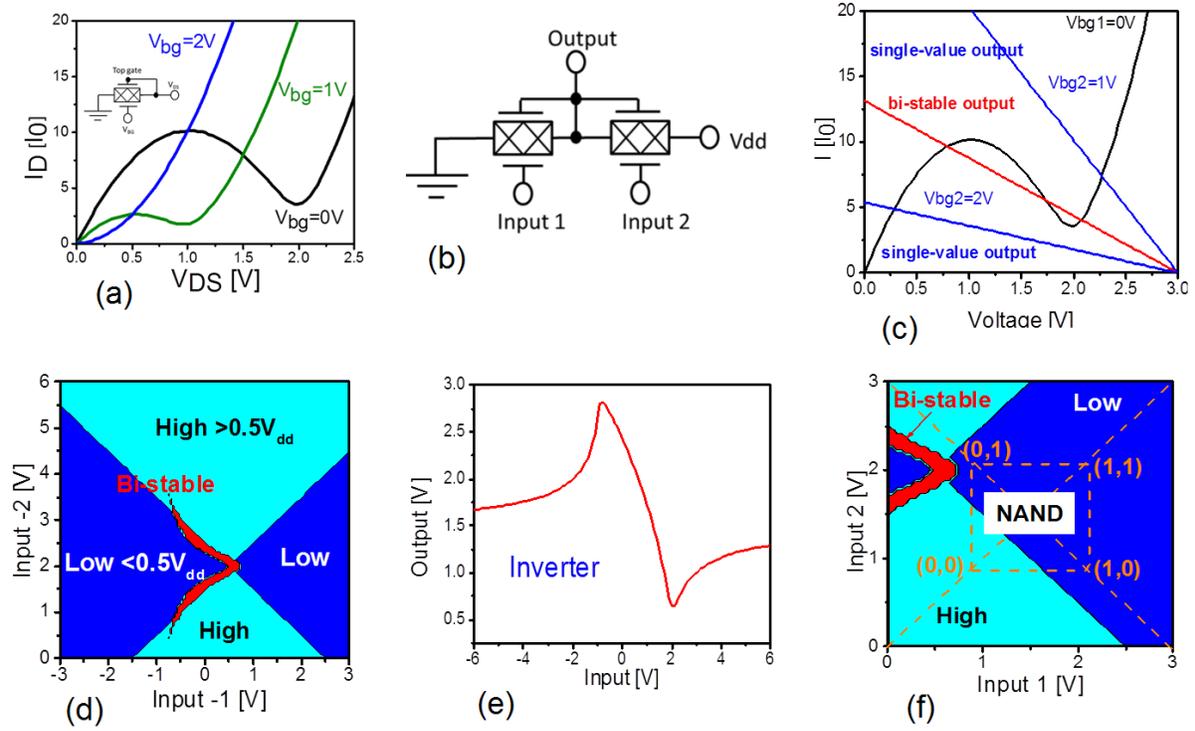

Figure 4 of 5

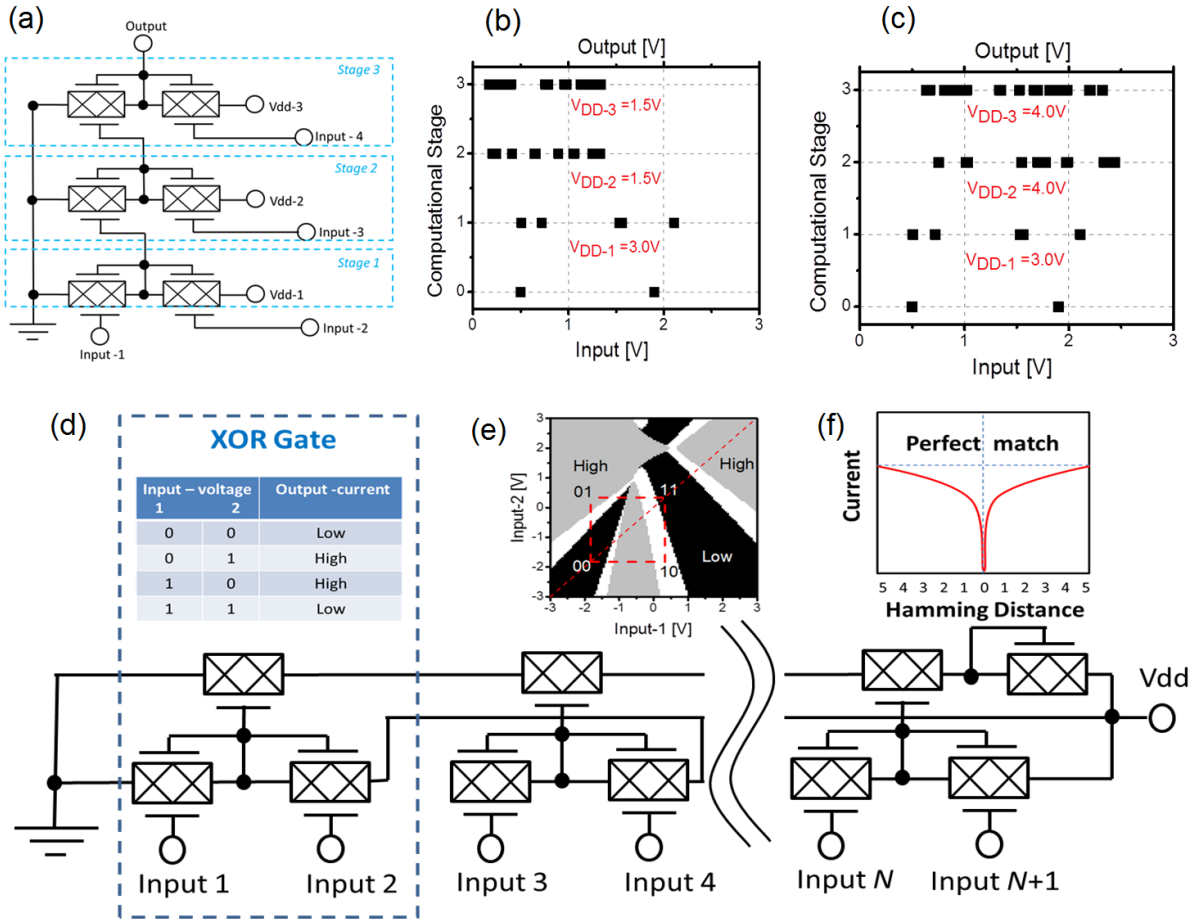

Figure 5 of 5